\documentclass[runningheads]{llncs}
\usepackage[T1]{fontenc}
\usepackage{cite}
\usepackage{subcaption}
\usepackage{amsmath,amssymb,amsfonts}
\usepackage{algorithmic}
\usepackage{textcomp}
\usepackage{comment}
\usepackage{xcolor}
\usepackage{float}
\def\BibTeX{{\rm B\kern-.05em{\sc i\kern-.025em b}\kern-.08em
		T\kern-.1667em\lower.7ex\hbox{E}\kern-.125emX}}
\newcommand\norm[1]{\left\lVert#1\right\rVert}
\newcommand{\Lapl}{\mathbf{\mathop{\mathcal{L}}}}
\newcommand{\Trans}[1]{{#1}^{\top}}

\newcommand{\Mat}[1]{\mathbf{#1}}

\newcommand{\Space}[1]{\mathbb{#1}}
\newcommand{\Set}[1]{\mathcal{#1}}

\newcommand{\BlockMatSquare}[4]{\left[\begin{matrix}#1 & #2\\#3 & #4\end{matrix}\right]}
\usepackage{adjustbox}
\usepackage{graphicx}
\begin{document}
\title{Neural Graph Collaborative Filtering Using Variational Inference}
%
%
\author{Narges Sadat Fazeli Dehkordi\inst{1} 
\and
Hadi Zare\inst{1} 
\and
Parham Moradi\inst{2} 
\and
Mahdi Jalili\inst{2}}
\authorrunning{N. Fazeli Dehkordi et al.}
%
\institute{Faculty of New Sciences and Technologies, University of Tehran, Tehran, Iran
\email{narges.fazeli.de@ut.ac.ir}\\
\email{h.zare@ut.ac.ir}\\
\and
School of Engineering, RMIT University, Melbourne, Australia\\
\email{parham.moradi@rmit.edu.au}\\
\email{mahdi.jalili@rmit.edu.au}}
\maketitle              
\begin{abstract}
The customization of recommended content to users holds significant importance in enhancing user experiences across a wide spectrum of applications such as e-commerce, music, and shopping.
Graph-based methods have achieved considerable performance by capturing user-item interactions. However, these methods tend to utilize randomly constructed embeddings in the dataset used for training the recommender, which lacks any user preferences. Here, we propose the concept of \textit{variational embeddings} as a means of pre-training the recommender system to improve the feature propagation through the layers of graph convolutional networks (GCNs).
The graph variational embedding collaborative filtering (GVECF) is introduced as a novel framework to incorporate representations learned through a variational graph auto-encoder which are embedded into a GCN-based collaborative filtering. This approach effectively transforms latent high-order user-item interactions into more trainable vectors, ultimately resulting in better performance in terms of recall and normalized discounted cumulative gain(NDCG) metrics. The experiments conducted on benchmark datasets demonstrate that our proposed method achieves up to 13.78\% improvement in the recall over the test data.


\end{abstract}
\section{Introduction}

Recommender systems can learn users' preferences based on their past interactions with items and suggest the most suitable options.
Collaborative filtering(CF) is one of the most popular approaches in recommender systems to identify user preferences\cite{17}.
This method operates under the assumption that users who have similar tastes in the past will continue to act similarly in the future. The development of deep neural networks has allowed researchers to enhance collaborative filtering in various ways. However, significant issues such as cold-start, scalability, and sparse data are still prevalent during the training phase\cite{17}.

To address the challenges of the cold-start and sparse data in CF, representation vectors, also called embeddings, were introduced\cite{b4,48}. These vectors are particularly useful in recommender systems to model the user preferences either explicitly, such as considering the user ratings\cite{41}, or implicitly\cite{b8}. 
A variety of embedding-based methods are introduced such as matrix factorization, neural network approaches, and auto-encoder methods \cite{43,b4,48}. Variational auto-encoders (VAEs) are specifically useful for the cold-start problem due to their probabilistic nature \cite{48}. This concept was integrated well into graph convolutional networks (GCNs) due to the graph nature of user-item interaction data\cite{b6}. 
Additionally, GCNs serve as a potent tool for capturing intricate relationships in graphs during collaborative filtering\cite{b6,b8}. Through information aggregation from neighboring nodes, GCNs propagate relevant features across the graph, facilitating a comprehensive understanding of user-item interactions. 
However, the initial embedding process often fails to consider user preferences comprehensively. 

In this article, we present a neural graph structured collaborative filtering method that exploits the capabilities of VGAE to feed the graph neural network-based recommender systems with latent embeddings of user-item interactions. A graph-based collaborative filtering method called GVECF, are presented based on  integration of variational auto-encoders and neural graph network structure to deal with the sparse data and cold-start issues. Notably, the encoded interaction matrix is fed into the NGCF as an initial embedding. This approach improves the performance of the recommender system in terms of recall, normalized discounted cumulative gain (NDCG)\cite{b4}, and other evaluation metrics. Specifically, the attained results reveal that the embeddings generated by VAEs will enhance the accuracy of the recommender system in comparison to the other integrated methods on neural graph collaborative filtering approach\cite{b8}.
The contributions of this work are:
\begin{itemize}
    \item We propose the concept of variational embeddings as structure-informed low-dimensionalized arrays in the latent space of a graph variational auto-encoder (GVAE). 
    \item We present a graph-based collaborative filtering method called GVECF, which takes the variational embeddings as pre-trained user preferences.
    \item The performance of GVECF is evaluated over several benchmark datasets.
\end{itemize}
\section{Related Works} 
\begin{itemize}
\item \textbf{Model-Based Collaborative Filtering}
Model-based CF has been widely explored as a powerful approach for recommendation systems\cite{23}. One popular technique is matrix factorization, which decomposes the user-item interaction matrix into low-rank matrices to capture latent factors\cite{24}. 
However, with the advent of deep learning\cite{27}, some algorithms are proposed by considering the idea of CF, including EnsVAE and NGCF\cite{48,b8}. 
\item \textbf{Variational Autoencoder Collaborative Filtering}
Variational autoencoders (VAEs) have gained significant attention in the field of recommendation systems due to their ability to learn non-linear latent representations and capture complex user-item interactions\cite{48}. 
By leveraging the reparametrization trick, VAEs can infer user preferences from observed ratings and generate personalized recommendations by sampling from the inferred latent space.
Some studies have proposed novel approaches to enhance VAE-based CF\cite{51,48}. One notable advancement is the incorporation of side information, such as user attributes and item features, into the VAE framework \cite{34}.
\item \textbf{Neural Graph Collaborative Filtering}
Neural graph collaborative filtering (NGCF) has recently emerged as a promising direction for recommendation systems\cite{b8}. This approach leverages graph-based representations to model the relationships between users and items. By treating users and items as nodes in a graph and exploiting GCNs\cite{b6}, it becomes possible to capture intricate network structures and propagate information through the graph. Furthermore, recent research has explored the integration of contrastive learning\cite{36},heterogeneous graph \cite{37}, and knowledge graphs\cite{38} to enhance NGCF-based models.
\end{itemize}
\section{Methoodology}
This section introduces the proposed GVECF model, and its architecture depicted in Fig. \ref{fig:1}. The framework consists of three components: (1) The user-item interaction matrix is pre-trained by encoding it into low-dimensional variational embeddings (VEs). (2) Several embedding propagation layers, and (3) the prediction layer to combine the refined embeddings and generate the affinity score of a user-item pair.

\subsection{Model Framework.}
\begin{figure}[h]
	\begin{center}
		\includegraphics[width=0.9\textwidth]{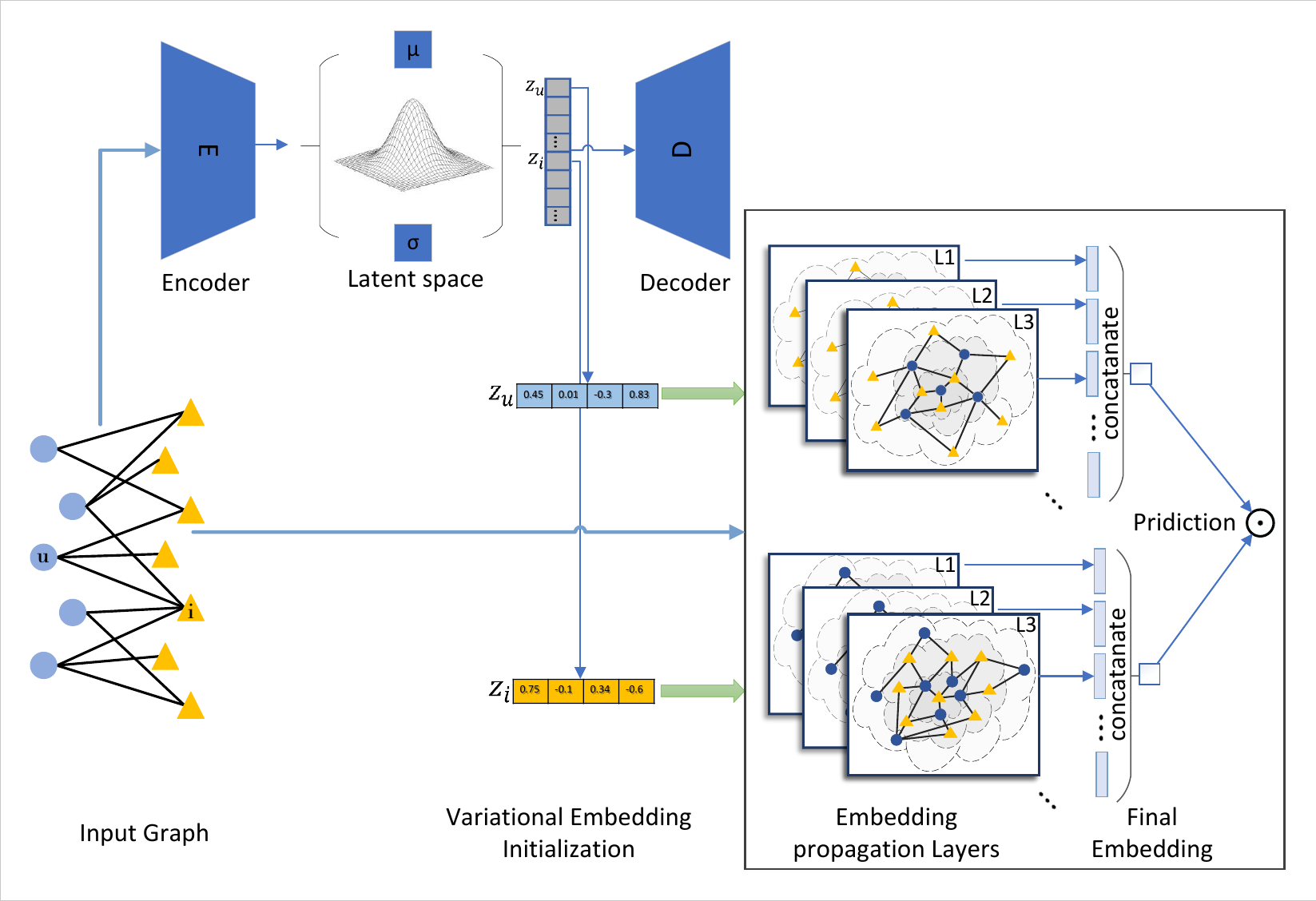}
		\caption{Architecture of the GVECF method}
		\label{fig:1}
	\end{center}
\end{figure}
\vspace{-5px}
We will elaborate on the GVECF in detail below, as presented in Fig. \ref{fig:1}.
\subsection*{Variational Embedding}

Assume there are N users represented as 
$u=\{u_{1},u_{2},\cdots,u_{N}\}$ and M items represented as $i=\{i_{1},i_{2},\cdots,i_{M}\}$ in a common recommendation context. Like previous graph-based studies\cite{b8}, this work specifically considers implicit feedback such as clicks or browsing. 
The user-item interaction matrix $R \in \mathbb{R}^{N\times M}$ can then be defined where the entry $R_{ui}$ equals 1 if the interaction (u,i) exists, and equals zero otherwise.


Assuming a bipartite graph model for the user-item interactions, the corresponding adjacency matrix becomes
\begin{align}
	\Mat{A}=\BlockMatSquare{\Mat{0}}{\Mat{R}}{\Trans{\Mat{R}}}{\Mat{0}},
\end{align}
the normalized graph Laplacian matrix is derived from $\Mat{A}$ as $\Lapl=\Mat{D}^{-\frac{1}{2}}\Mat{A}\Mat{D}^{-\frac{1}{2}}$, 
where $\Mat{D}$ is the diagonal degree matrix whose $t$-th diagonal element is $D_{tt}=|\Set{N}_{t}|$ with $\Set{N}_{t}$ being the set of 1st-hop neighbors of the node $t$. As such, the nonzero off-diagonal entries are $\Lapl_{ui}=\frac{1}{\sqrt{|\Set{N}_{u}||\Set{N}_{i}|}}$.
In the previous approaches to collaborative filtering recommendation that were based on graph neural networks (GNNs)\cite{b8}, researchers used initializer functions, such as Xavier initializer\cite{55},  
to generate the initial embeddings of users and items which are then propagated through the layers of GNN. However, the user-item interactions have played no part in constructing the embeddings prior to propagation. In the proposed model, we follow this idea by utilizing VEs obtained from the hidden layers of a VGAE\cite{b7} as a mechanism to pre-train the dataset.
While maintaining the user-item graph connectivity, the VGAE embeds the high-order interactions into pre-trained low-dimensional representations that will provide a head-start in the training phase, as substantiated in the results section. 
a brief overview of VGAE is given by considering two main parts, Inference and Generating models\cite{51}.  
\subsubsection{Inference model}
The objective of an inference model is to represent the original input data as a probability distribution. The encoder utilizes a GNN to estimate the mean and variance in latent user preference distribution. To achieve this, distribution $q_{\phi} (\mathbf{z}_{u} |\mathbf{x}_{u} )$ along with a Gaussian prior distribution $p(\mathbf{z}_{u})=\mathcal{N}(\mathbf{z}_{u}|0,\Mat{I})$ is utilized to estimate the often inaccessible posterior distribution $p(\mathbf{z}_{u} |\mathbf{x}_{u} )$, where $\mathbf{x}_{u}$ and $\mathbf{z}_{u}$ represent the original interaction vector and the d-dimensional hidden vector of user u, respectively\cite{b8}:
\begin{equation}
	q(\mathbf{Z}\,|\,\mathbf{X},\mathbf{A}) = \prod_{i=1}^M q(\mathbf{z}_i\,|\,\mathbf{X},\mathbf{A})\prod_{u=1}^N q(\mathbf{z}_u\,|\,\mathbf{X},\mathbf{A})\,
\end{equation}
where for user u, the posterior distribution of $\mathbf{z}_u$ is constructed with mean vector $\boldsymbol{\mu}_u$ and standard deviation $\boldsymbol{\sigma}_u$:
\begin{equation}\label{eq:mu}
		q(\mathbf{z}_n\,|\,\mathbf{X},\mathbf{A})  = \mathcal{N}(\mathbf{z}_n\,|\, \boldsymbol{\mu}_n, \mathrm{diag}(\boldsymbol{\sigma}_n^2))\,\,\quad n\in\{u,i\}
\end{equation}
In equation \ref{eq:mu}, $\boldsymbol{\mu} = \mathrm{GCN}_{\boldsymbol{\mu}}(\mathbf{X}, \mathbf{A})$ is the matrix of mean vectors $\boldsymbol{\mu}_i$ and $\log\boldsymbol{\sigma} = \mathrm{GCN}_{\boldsymbol{\sigma}}(\mathbf{X}, \mathbf{A})$ is the variance vector.
\subsubsection{Generative Model}
Once the distribution vectors $\mathbf{z}_u$ for each user node in the graph are acquired, the goal is to utilize a generative model to recreate the graph representing the original interactions between users and items. This method involves reconstructing the adjacency matrix A of the input graph by employing inner product calculations between implicit variables.
\begin{equation}
		p\left(\mathbf{A\,|\,\mathbf{Z}}\right) = \prod_{i=1}^M\prod_{j=1}^N p\left(A_{ij}\,|\,\mathbf{z}_i,\mathbf{z}_j\right)\,\,\,\,\text{where}\,\, p\left(A_{ij}=1\,|\,\mathbf{z}_i,\mathbf{z}_j\right) = \sigma(\mathbf{z}_i^\top\mathbf{z}_j) \, 
\end{equation}
where $A_{ij}$ are the elements of $\mathbf{A}$ and $\sigma(\cdot)$ is the logistic sigmoid function. The output of the generative model is used to train the GVAE model.
\subsection{Embedding Propagation}
After encoding the user-item interactions to low-dimensional vectors which have successfully embodied important features within their structure, we use them as initial embeddings in the message-passing architecture of a GNN \cite{b6}. This architecture can capture the interactions in any distance and improve the embedding of users and items. In order to explain the architecture we first introduce the data propagation mechanisms through consecutive layers: message construction and message aggregation. 
\subsubsection{Message Construction}
Once the VGAE outputs the latent embeddings $\Mat{z}_u$ and $\Mat{z}_i$, these vectors are fed into a GNN as initial embeddings, with the ultimate goal of recommending similar items to the users of similar preferences. The first step to train this GNN involves creating the message that will be propagated through its layers.
If there exists an edge between user u and item i in a bipartite graph, this edge is considered as a message from item i to user u which is constructed as follows:
\begin{gather}\label{equ:0-message}
	\Mat{m}_{u\leftarrow i}=f(\Mat{z}_{i},\Mat{z}_{u},p_{ui})=\frac{1}{\sqrt{|\Set{N}_{u}||\Set{N}_{i}|}}\Big(\Mat{W}_{1}\Mat{z}_{i} + \Mat{W}_{2}(\Mat{z}_{i}\odot\Mat{z}_{u})\Big)
\end{gather}
Here $\Mat{m}_{u\leftarrow i}$ is the message embedding. $f(\cdot)$ is the message encoding function that takes VEs $\Mat{z}_{i}$ and $\Mat{z}_{u}$ as input, and uses the coefficient $p_{ui}$ to control the decay factor on each propagation on edge $(u,i)$.
By enhancing the representations through the incorporation of first-order connectivity modeling, adding additional layers of embedding propagation allows for a deeper exploration of higher-order connectivity details. These extended connections play a pivotal role in capturing the collaborative cues necessary for estimating the relevance score between users and items. For the $l^{th}$ layer, the message construction function can be generalized as follows\cite{b8}:
\begin{gather}\label{equ:l-message}	
	\Mat{m}^{(l)}_{u\leftarrow i}=p_{ui}\Big(\Mat{W}^{(l)}_{1}\Mat{z}^{(l-1)}_{i} + \Mat{W}^{(l)}_{2}(\Mat{z}^{(l-1)}_{i}\odot\Mat{z}^{(l-1)}_{u})\Big),\\
	\Mat{m}^{(l)}_{u\leftarrow u}=\Mat{W}^{(l)}_{1}\Mat{z}^{(l-1)}_{u}
\end{gather}
where a self-message is also defined for each user in second line of equation \ref{equ:l-message} in order to include its embedding history in the training process. It should be noted that a message construction procedure similar to equation \ref{equ:l-message} is established for each item $i$.
\subsubsection{Massage Aggregation} 
To aggregate the constructed messages after the first embedding propagation layer for user $u$, the following function is defined.
\begin{gather}\label{equ:1-aggregator}
	\Mat{z}^{(1)}_{u}=\text{LeakyReLU}\Big(\Mat{m}_{u \leftarrow u} + \sum_{i\in\Set{N}_{u}}\Mat{m}_{u\leftarrow i}\Big),
\end{gather}
for messages propagated from neighbors $\Set{N}_u$, we consider the self-connection of $u$ which retains the information of original features ($\Mat{W}_1$ is the weight matrix shared with the one used in Equation (\ref{equ:0-message})).
\begin{gather}\label{equ:1-aggregator}
	\Mat{m}_{u\leftarrow u}=\Mat{W}_1\Mat{z}_{u},
\end{gather}
Similar to \cite{b8}, the LeakyReLU activation function is used to alleviate the vanishing gradient problem. In order to capture the $l$-hop connections of users and items, we stack up $l$ layers of graph convolutional layers and generalize the message augmentation mechanism. Writing in the matrix form, message augmentation is defined as follows\cite{b6}:
\begin{equation}\label{equ:rule}
	\resizebox{0.85\hsize}{!}{$\Mat{Z}^{(l)}=\text{LeakyReLU}\Big((\Lapl+\Mat{I})\Mat{Z}^{(l-1)}\Mat{W}^{(l)}_{1} + \Lapl\Mat{Z}^{(l-1)}\odot\Mat{Z}^{(l-1)}\Mat{W}^{(l)}_{2}\Big)$}
\end{equation}
In equation \ref{equ:rule}, $\Mat{Z}^{(l)}\in\Space{R}^{(N+M)\times d_{l}}$ are the representations for users and items after $l$ steps of embedding propagation. 
\subsubsection{Model Prediction}
According to \cite{b8}, each layer of the GNN outputs latent learned VEs for user $u$ and item $i$, which we denote by $\{\Mat{z}^{(1)}_{u}$ and $\{\Mat{z}^{(1)}_{i},\cdots,\Mat{z}^{(L)}_{i}\}$ respectively. To explicitly include the embeddings associated with each layer, they are concatenated to generate the final representation,
\begin{equation}\label{equ:final-rep}
	\Mat{e}^{*}_{u}= \Mat{e}^{(0)}_{u}\Vert\cdots\Vert\Mat{e}^{(L)}_{u}, \quad
	\Mat{e}^{*}_{i}=\Mat{e}^{(0)}_{i}\Vert\cdots\Vert\Mat{e}^{(L)}_{i},
\end{equation}
where $\Vert$ is the concatenation operation.
An inner product is applied to quantify how much the user $u$ prefers the item $i$: 
\begin{align}
	\hat{y}_{\text{NGCF}}(u,i)=\Trans{\Mat{e}^{*}_{u}}\Mat{e}^{*}_{i}.
\end{align}
\section{Experiments}
\subsection{\textbf{Test Configuration}}
\subsubsection{\textbf{Dataset} }\label{AA}
Four benchmark datasets are considered: Gowalla\cite{20}, Yelp\cite{21}, and Amazon-Books\cite{22}. In all cases, we consider the datasets as implicit feedback. A summary of each dataset is presented in Table \ref{tab:dataset}.
\begin{table}[t]
	\caption{Statistics of the datasets.}
	\label{tab:dataset}
	\begin{center}
		\resizebox{0.7\textwidth}{!}{
			\begin{tabular}{l|r|r|r|r}
				\hline
				\multicolumn{1}{c|}{Dataset} & \multicolumn{1}{c|}{\#Users} & \multicolumn{1}{c|}{\#Items} & \multicolumn{1}{c|}{\#Interactions} & \multicolumn{1}{c}{Density} \\ \hline\hline
				Gowalla & $29,858$ & $40,981$ & $1,027,370$ & $0.00084$ \\ \hline
				Yelp$^{*}$ & $31,668$ & $38,048$ & $1,561,406$ & $0.00130$ \\ \hline
				Amazon-Book & $52,643$ & $91,599$ & $2,984,108$ & $0.00062$ \\ \hline
				
		\end{tabular}}
	\end{center}
		\vspace{-15px}
\end{table}\newline
\textbf{Gowalla:} This dataset is derived from Gowalla, a platform where users share their locations through check-ins. To ensure the dataset's quality, the 10-core setting is applied\cite{b8}.\\ 
\textbf{Yelp:} Adopted from the 2018 version of the Yelp challenge, local establishments such as restaurants and bars are considered as the items of interest in this dataset. To maintain the integrity of the data, we employ the identical 10-core setting.\\
\textbf{Amazon Book:} The Amazon-review dataset is derived from Amazon's product network. For our study, we specifically focus on the Amazon-book subset within this dataset. Likewise, we implement the 10-core configuration to uphold data quality standards.
\subsubsection{\textbf{Implementation}}
We used the PyTorch package to implement VGAE, GVECF, and NGCF. The codes are deployed on two systems equipped with NVIDIA GeForce RTX 3090 GPU and NVIDIA GeForce RTX 3070 GPU, with 24G and 8G of GPU memory. For VGAE, a 3-layer GNN is implemented with a learning rate 0.01. The batch and embed sizes of 1024 and 64 are deemed appropriate On GVECF and NGCF algorithms by considering memory limitations. A pairwise BPR loss is optimized in order to optimize all model weights using the same formulation provided in \cite{b8}. Moreover, a grid search algorithm is applied to tune the set of hyperparameters comprising the learning rate $lr$, the regularization strength $\lambda_{\text{reg}}$, and the node dropout ratio $nd$, which are tuned over $(lr,\lambda_{\text{reg}},nd)\in\{5e-4,1e-4,5e-5,1e-5,5e-6\}\times\{5e-4,1e-5,1e-6\}\times\{0.1,0.2\}$, while the message dropout ratio was fixed at 0.1 for reduced computational costs.

\subsubsection{\textbf{Baseline}}

To showcase its efficacy, we conduct a comparative analysis of the suggested GVECF model against these state-of-the-art methods,

\begin{itemize}
	\item MF\cite{b9}: This technique involves the process of matrix factorization, which has been enhanced through the integration of the Bayesian personalized ranking (BPR) loss. 
	\item NCF\cite{b4}: Neural Collaborative Filtering is a recommendation system that combines neural networks with collaborative filtering. It employs multiple hidden layers in a two-layer neural network architecture.
	\item HOP-REC\cite{b5}: HOP-Rec presents an interesting approach that merges factorization and graph-based models.
	\item NGCF\cite{b8}: Neural Graph Collaborative Filtering is a state-of-the-art recommendation model that employs GCNs to effectively capture collaborative filtering signals between users and items. 	
\end{itemize}
	  \begin{table}[t]
		\begin{center}
			\caption{Overall Performance Comparison}
			\label{tab2}
			\resizebox{0.7\textwidth}{!}{
				\begin{tabular}{|c|c c|c c|c c|c c|}
					\hline
					& \multicolumn{2}{c|}{Gowalla} & \multicolumn{2}{c|}{Yelp$^{}$} & \multicolumn{2}{c|}{Amazon-Book}                              \\ \hline
					& Recall & NDCG & Recall & NDCG & Recall & NDCG  \\ \hline
					{MF}      & 0.1291 & 0.1109 & 0.0433 & 0.0354 & 0.0250 & 0.0196  \\ \hline
					{NCF}   & 0.1399 & 0.1212 & 0.0451 & 0.0363 & 0.0258 & 0.0200  \\ \hline
					
					{HOP-Rec} & 0.1399 & 0.1214 & 0.0517 & 0.0428 & 0.0309 & 0.0232 \\ \hline
					
					{NGCF} & ${0.1555^{}}$ & ${0.1324^{}}$ & ${0.0548^{}}$ & ${0.0448^{}}$ & ${0.03155^{}}$ & ${0.02414^{}}$  \\ \hline
					\textbf{GVECF} & ${0.1578^{*}}$ & ${0.1353^{*}}$ & ${0.0566^{*}}$ & ${0.0463^{*}}$ & ${0.0359^{*}}$ & ${0.0273^{*}}$ \\ \hline
					\%Improv. & \textbf{1.47\%} & \textbf{2.22\%} & \textbf{3.15\%} & \textbf{3.28\%} & \textbf{13.78\%} & \textbf{13.09\%} \\ \hline
			\end{tabular}}
		\end{center}
	 	\vspace{-10px}
\end{table}

\subsubsection{Evaluation Metrics}
In the test set, we consider un-interacted items for each user as negative items. Subsequently, each technique generates preference scores for all items except those designated as positive items in the training set. The evaluation of top-K recommendation and preference ranking employs established assessment protocols, namely recall@K, and NDCG@K, with a default setting of K = 20. The metrics' averages are computed across all users in the test set, serving as a measure of the methods' effectiveness \cite{b4,b8,b5}.

\subsection{Results comparison}
The results obtained from the proposed model and several baseline methods\cite{b4,b8,b5} are presented in table \ref{tab2}. Our findings indicate that GVECF exhibits superior recall and NDCG for top-20 recommendation, compared to other methods for all benchmark datasets provided in \ref{tab2}. The proposed method not only enhances recommendation capability due to the pre-trained embeddings of VGAE but also exhibits better convergence and computational efficiency.

\subsection{Study of GVECF}
We first study the accuracy and time performance of GVECF over recall and NDCG metrics as the recommender is trained through 400 epochs, and compare the results with NGCF. Then, we study the effects of embed size on GVECF compared with NGCF. In order to demonstrate the effectiveness of the variational formulation for low-density datasets, a graph auto-encoder with similar settings to our VGAE pre-trainer is also used to construct embeddings at the initial stage. 
\subsubsection{Effect of Variational Embedding}

\begin{figure}[htbp]
	\begin{subfigure}[b]{0.3\textwidth}
	\centering
 \resizebox{\linewidth}{!}{
	\includegraphics[width=\textwidth]{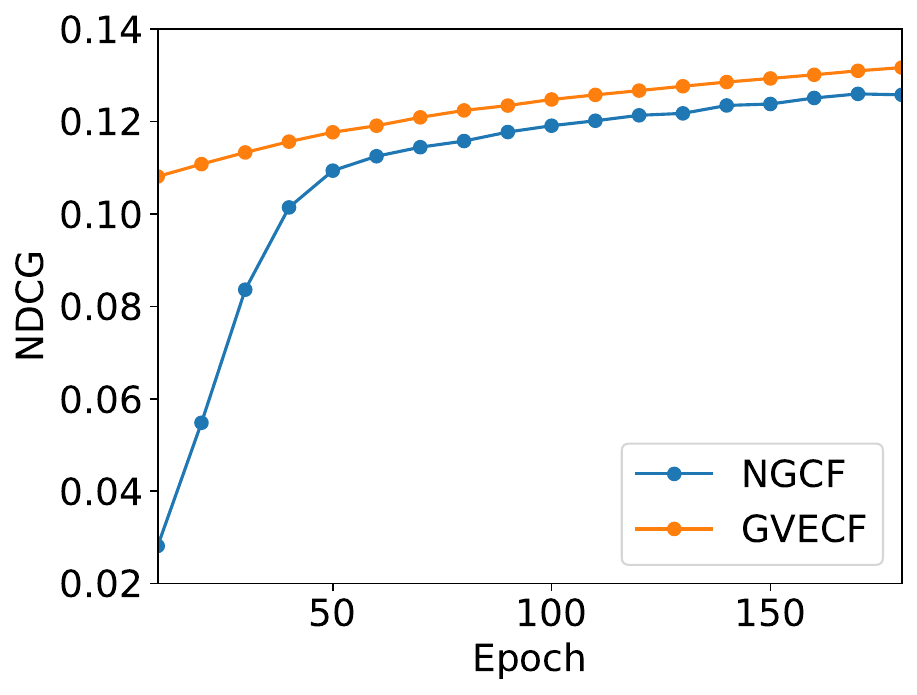}
}
	\caption{}
	\label{fig:3}
    \end{subfigure}
	\begin{subfigure}[b]{0.3\textwidth}
	\centering
 \resizebox{\linewidth}{!}{
	\includegraphics[width=\textwidth]{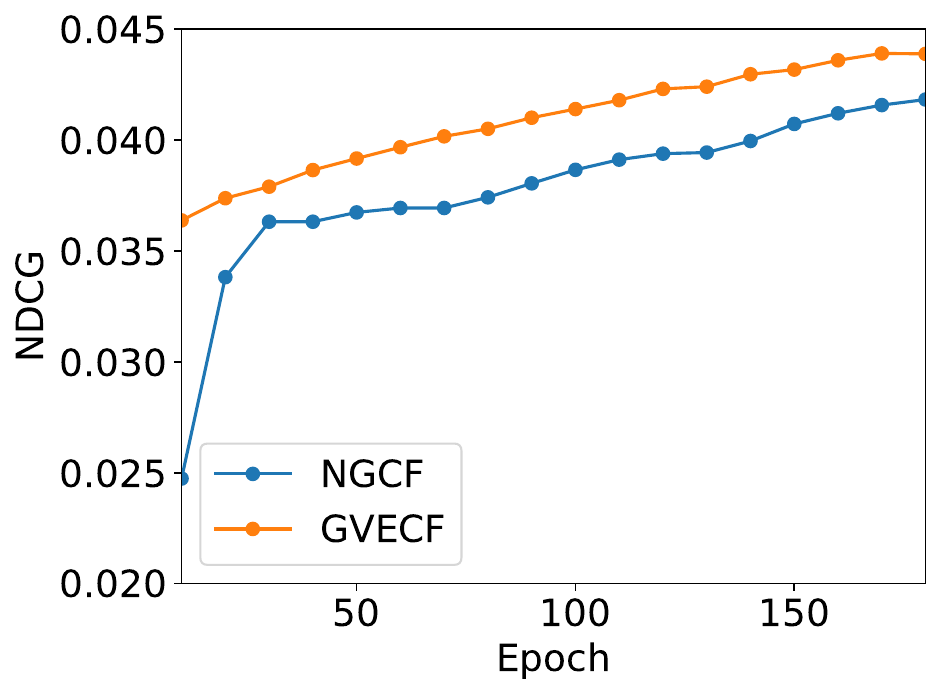}}
	\caption{}
	\label{fig:4}
	\end{subfigure}
	\begin{subfigure}[b]{0.3\textwidth}
	\centering
 \resizebox{\linewidth}{!}{
	\includegraphics[width=\textwidth]{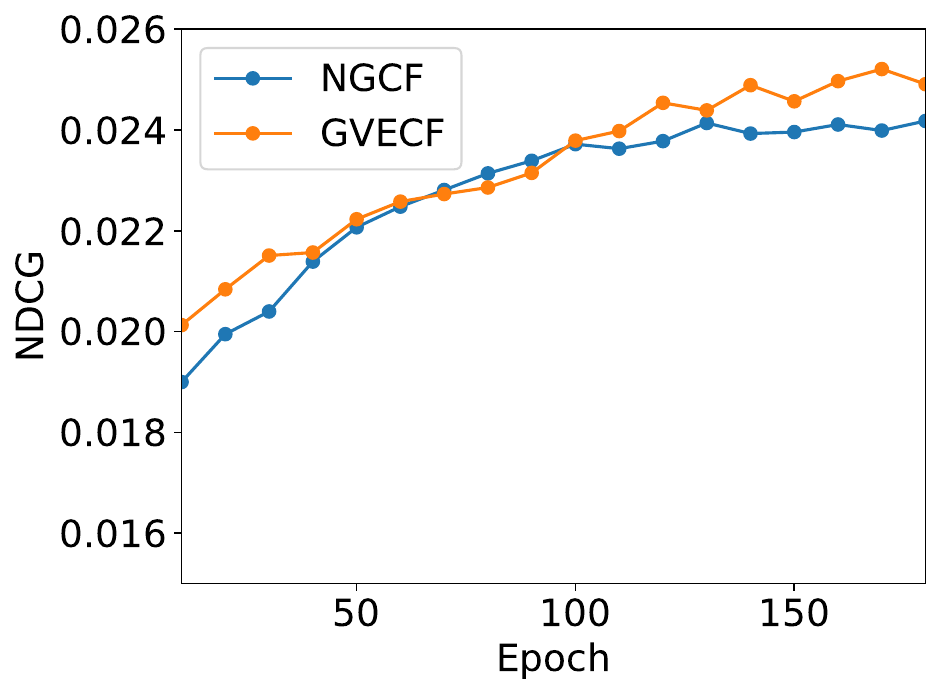}}
	\caption{}
	\label{fig:5}
	\end{subfigure}
\caption{(a) NDCG on Gowalla, (b) NDCG on Yelp (c) NDCG on Amazon-Book}
\label{fig:3tot}
\end{figure}

Fig. \ref{fig:3tot} illustrates the performance of GVECF and NGCF on NDCG metric over 180 epochs. We have computed NDCG in 10-epoch cycles, as represented with filled dots in Fig. \ref{fig:3tot}. GVECF shows a substantial head-start in Gowalla and Yelp datasets due to pre-training (47.0\% and 7.6\% improvement over NGCF, respectively). The performance of both methods is close for the Amazon dataset in the initial epochs; however, an eventual improvement of $13.09\%$ and $13.78\%$ are achieved for NDCG and recall metrics, which are the most remarkable among the three datasets. Considering the dataset characteristics provided in table \ref{tab:dataset}, our VE approach has ameliorated the sparsity problem from which the Amazon dataset particularly suffers.


\vspace{-25px}
\subsubsection{Effect of Embed Size}
The initial embed size has a crucial role in capturing messages in VE, as well as in embedding propagation. The larger the embedding size, the more features that are examined, and subsequently, the more accurate recommendations the model provides. If the embedding dimension is too small, it limits the representation capacity of embedding vectors, preventing them from effectively utilizing the information present in the user-item graph. As depicted in Fig. \ref{fig:3}, On the other hand, the performance initially shows rapid enhancement as the embedding dimension increases, but it eventually plateaus and can even decline from the optimal point due to overfitting. Enlarging the size of embeddings from 16 to 64, the GVECF method enjoys an improvement of  12.8\%, 11.3\%, and 35.5\% for the Gowalla, Yelp, and Amazon book datasets. These numbers reduce to 1.1\%, 1.6\%, and 1.6\% if a further enlargement from 64 to 80 takes place. Accordingly, an embed size of 64 is deemed appropriate for all scenarios in the paper. 
\begin{figure}[t]
	\begin{subfigure}[b]{0.3\textwidth}
	\centering
	\includegraphics[width=\columnwidth]{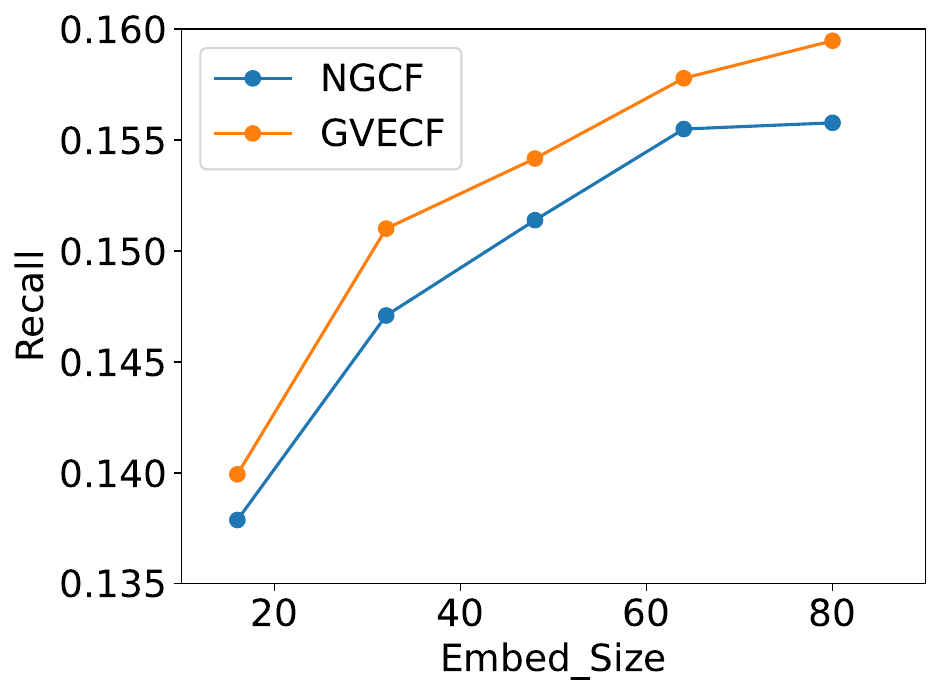}
	\caption{}
	\label{fig:16}
	\end{subfigure}
	\begin{subfigure}[b]{0.3\textwidth}
	\centering
	\includegraphics[width=\columnwidth]{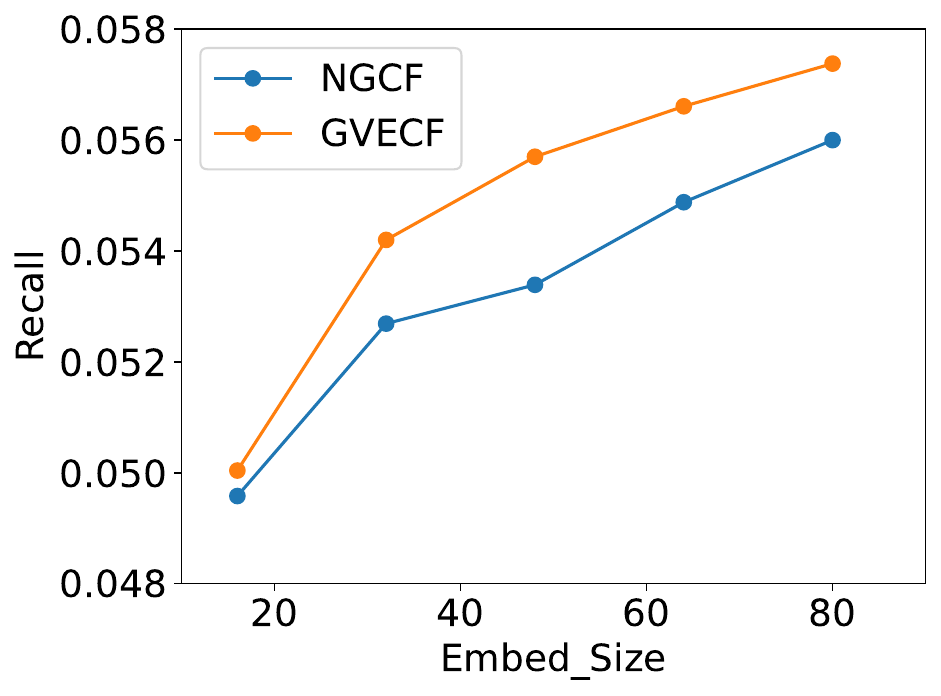}
	\caption{}
	\label{fig:17}
	\end{subfigure}
	\begin{subfigure}[b]{0.3\textwidth}
	\centering
	\includegraphics[width=\columnwidth]{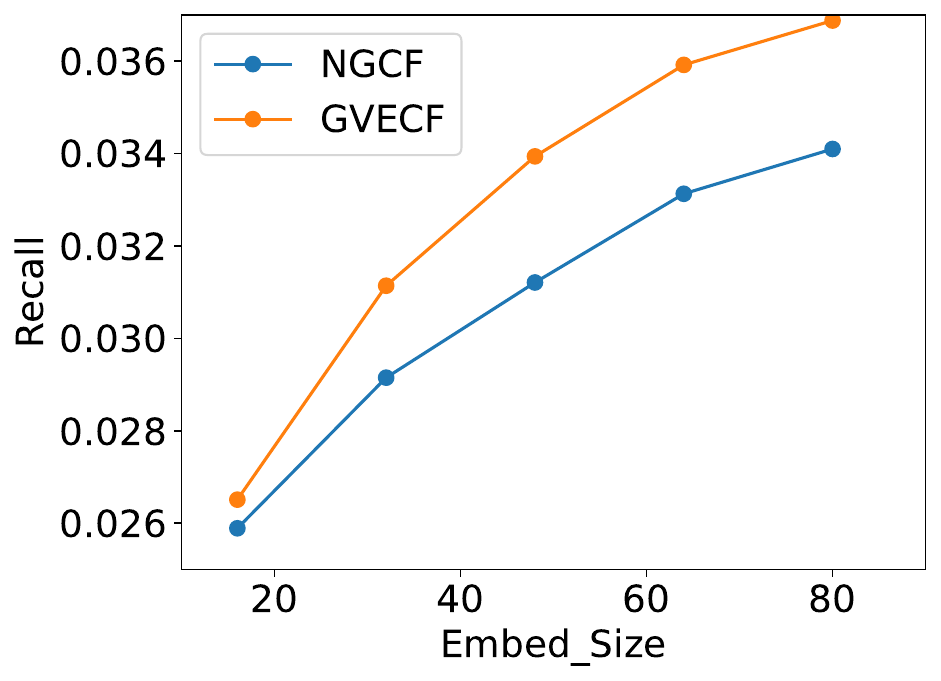}
	\caption{}
	\label{fig:18}
	\end{subfigure}
\caption{(a) effect of embed size on Gowalla, (b) effect of embed size on yelp,  (c) effect of embed size on Amazon-Book}
\label{fig:4tot}
\end{figure}
\vspace{-15px}
\subsubsection{Effect of initial embedding method}
The effect of pre-training on the recommender system was investigated in the previous section. The particular implication of a variational formulation for embedding construction is analyzed by comparing the results with a similar GNN-based recommender system whose initial embeddings are constructed with a graph auto-encoder (GAE) \cite{b7} that shares the same structure with the VGAE used for the construction of our VEs. According to Table\ref{tab3}, VAE yields better recall and NDCG for all datasets. Notably, the GVECF pre-trained with the VGAE approach only slightly outperforms the rival GVECF trained on simple graph auto-encoded embeddings for Gowalla and Yelp datasets, whereas both methods are approximately equally better than the baseline NGCF which uses Xavier initialization. This observation suggests that structure-informed embeddings can positively affect the training of the GNN-based recommender regardless of their variational characteristics. The advantages of VE are better demonstrated in the right end column of table \ref{tab3}, where GVECF\_VGAE improves the baseline NGCF and GVEC\_GAE by 13.7\% and 12.6\% over the recall metric. Our proposed method is a viable approach for extremely low-density scenarios since the scarce user preferences are replaced with approximate embeddings sampled from a tuned probability distribution.
\begin{table*}[t]
	\begin{center}
		\caption{Overall Performance Comparison.}
		\label{tab3}
		\resizebox{0.9\textwidth}{!}{
			\begin{tabular}{l|c c|c c|c c}
				\hline
				& \multicolumn{2}{c|}{Gowalla} & \multicolumn{2}{c|}{Yelp} & \multicolumn{2}{c}{Amazon-Book} \\ 
				& Recall & NDCG & Recall & NDCG & Recall & NDCG  \\  \hline\hline
				
				
				{NGCF} & ${0.1555^{}}$ & ${0.13240^{}}$ & ${0.05488^{}}$ & ${0.04481^{}}$ & ${0.03155^{}}$ & ${0.02414^{}}$  \\ \hline 
				{GVECF\_{GAE}} & ${0.15705^{}}$ & ${0.13426^{}}$ & ${0.05641^{}}$ & ${0.04585^{}}$ & ${0.03181^{}}$ & ${0.02443^{}}$  \\ \hline
				
				\textbf{{GVECF\_{VGAE}}} & $\boldsymbol{0.15779^{*}}$ & $\boldsymbol{0.13534^{*}}$ & $\boldsymbol{0.05661^{*}}$ & $\boldsymbol{0.04628^{*}}$ & $\boldsymbol{0.0359^{*}}$ & $\boldsymbol{0.0273^{*}}$  \\ \hline
		\end{tabular}}
		
	\end{center}
\end{table*}
\subsubsection{Computational Effort}

\begin{figure}[t]
	\begin{subfigure}[b]{0.3\textwidth}
	\centering
	\includegraphics[width=\textwidth]{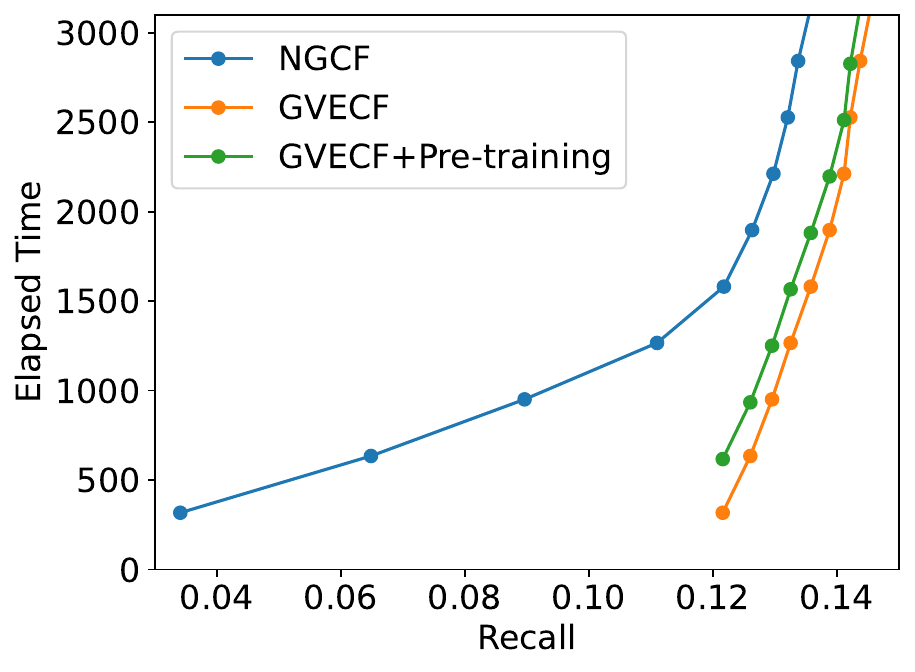}
	\caption{}
	\label{fig:6}
    \end{subfigure}
    \begin{subfigure}[b]{0.313\textwidth}
    \centering
	\includegraphics[width=\textwidth]{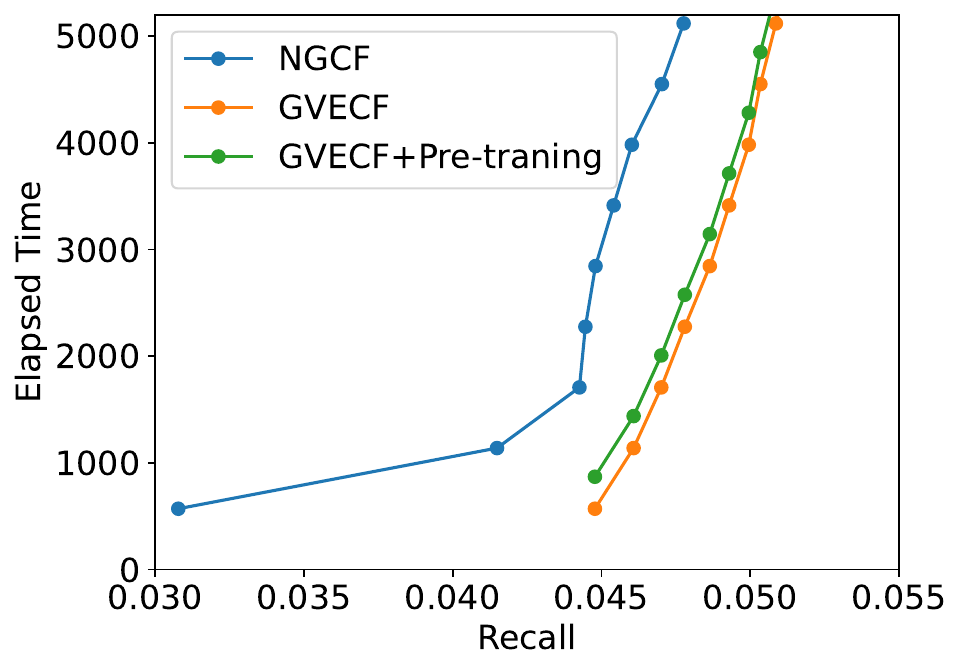}
	\caption{}
	\label{fig:7}
    \end{subfigure}
    \begin{subfigure}[b]{0.3\textwidth}
	\centering
	\includegraphics[width=\textwidth]{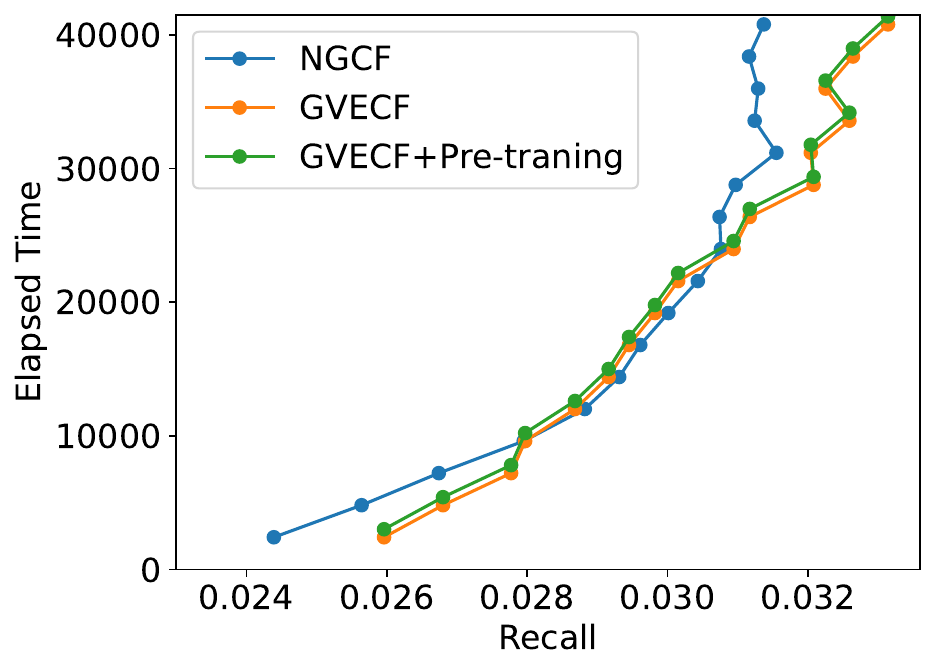}
	\caption{}
	\label{fig:8}
	\end{subfigure}
\caption{(a) training time on Gowalla, (b) training time on Yelp,  (c) training time on Amazon-Book}
\label{fig:5tot}
\end{figure}

Fig. \ref{fig:5tot} depicts the time performance of GVECF and NGCF methods. For a given elapsed time, the recalls obtained by the GVECF are constantly higher compared with the results of NGCF in the Gowalla and Yelp datasets. The average time elapsed for each epoch is nearly equal for the NGCF and GVECF approaches. NGCF reaches its best results in epochs 400, 400, and 130 for the Gowalla, Yelp, and Amazon datasets. Our proposed method reaches these target values at epochs 219, 209, and 115. Therefore, it can be inferred that GVECF reduced the computational effort of NGCF by 82.6\%, 91.4\%, and 13\% for the aforementioned datasets, with a possibility to acquire higher accuracies once convergence is achieved. 
\vspace{-20px}
\section{conclusion and future work}
In this paper, we proposed GVECF as a pre-training scheme in GNN-based CF methods. Our proposed method utilizes the ability of VGAEs to capture hidden features of possibly higher-order user-item interactions. Experiments on three benchmark datasets demonstrated the effectiveness of our
proposed method. The proposed method outperformed its competitors in terms of recall and NDCG. The recall is improved by 13.78\% for the extremely low-density Amazon dataset, further emphasizing the effectiveness of the proposed VE in such scenarios. GVECF maintains its superiority over the baseline NGCF for both small and large embed sizes while offering the same recall provided by NGCF with substantially lower computations.

\end{document}